\documentclass[%
 reprint,
 amsmath,amssymb,
 aps,
]{revtex4-2}
\usepackage{subcaption}
\usepackage{graphicx}
\usepackage{dcolumn}
\usepackage{bm}
\usepackage{ulem}
\usepackage{svg}
\begin{document}

\preprint{APS/123-QED}

\title{Compact and efficient quantum frequency conversion of a fiber-pigtailed single-photon source}

\author{Mathis Cohen$^{1}$, Anthony Martin$^{1}$, Romain Dalidet$^{1}$, Florian Pastier$^{2}$, Marie Billard$^{2}$, Aristide Lemaitre$^{3}$, Valérian Giesz$^{2}$, Niccolo Somaschi$^{2}$, Sarah Thomas$^{3}$, Pascale Senellart-Mardon$^{3}$, Sébastien Tanzilli$^{1}$, Laurent Labonté$^{1}$}
\email{laurent.labonte@univ-cotedazur.fr}
\affiliation{$^{1}$Université Côte d’Azur, CNRS, Institut de physique de Nice, France}
\affiliation{$^{2}$Quandela SAS, Palaiseau, France}
\affiliation{$^{3}$Université Paris-Saclay, CNRS, Centre de Nanosciences et de Nanotechnologies (C2N), Palaiseau, France}

\begin{abstract}
Quantum frequency converters are key enabling technologies in photonic quantum information science to bridge the gap between quantum emitters and telecom photons. Here, we report a coherent frequency converter scheme combining a fiber-coupled nonlinear optical Lithium Niobate waveguide with a fiber-pigtailed single-photon source based on semiconductor quantum dots. Single and indistinguishable photons are converted from $925.7$ nm to the telecommunication C-band,  with a $48.4$\% end-to-end efficiency and full preservation of single-photon purity and indistinguishability. The integration of the two fiber-based modules achieving top-level performance represents an important step toward the practical interconnection of future quantum information processing systems operating at different wavelengths.
\end{abstract}

 \maketitle
 
 \section{introduction}
 Distributing quantum resources over long-distance optical fiber networks is currently a challenge~\cite{labonte_integrated_2024}. This has already allowed quantum nonlocality to be demonstrated in a wide variety of applications, including device-independent quantum key distribution~\cite{PhysRevLett.98.230501, Pironio_2009}, deterministic quantum teleportation~\cite{nadlinger_experimental_2022}, entanglement-based clock synchronisation ~\cite{komar_quantum_2014, pelet_operational_2023}, distributed quantum sensors~\cite{chen_quantum_2022}, and quantum repeaters ~\cite{PhysRevLett.81.5932}. The latter approach appears to be a viable solution for transferring photonic quantum states from one location to another. A major challenge relies on the incompatibility in terms of the wavelength between telecom photons ($\sim1550~$nm), which represent natural flying qubit carriers for establishing entanglement, and deterministic single-photon emitters (nodes based on atoms, semiconductor quantum dots, or colour centres)  that mainly operate in the near-infrared (NIR) range ($\sim780-950~$nm). Quantum frequency conversion (QFC) addresses this challenge by bridging the gap between distinct spectral ranges by transferring the wavelength of a photonic state to another while preserving its quantum properties. QFC also presents an alternative solution for establishing true single-photon emission at telecommunication wavelengths, as the typical emission wavelength of mature single-photon emitters lies in the NIR range ($\sim900-950~$nm)~\cite{somaschi_near-optimal_2016}. As such, QFC has received a lot of attention over the past years through the demonstration of photonic quantum information downconverters~\cite{tanzilli_photonic_2005, lenhard_coherence_2017, Kaiser:19}, matter-light entanglement~\cite{de_greve_quantum-dot_2012, ikuta_polarization_2018, PhysRevLett.105.260502, yu_entanglement_2020, bock_high-fidelity_2018, tchebotareva_entanglement_2019, strassmann_spectral_2019, krutyanskiy_light-matter_2019}, quantum storage of entangled telecom-wavelength photons~\cite{saglamyurek_quantum_2015}, as well as bright and pure single-photon source at telecom wavelength~\cite{, zaske_visible--telecom_2012, Pelc:12, Kambs:16, da_lio_pure_2022,morrison_bright_2021, singh_quantum_2019}.\\
QFC relies on nonlinear processes, such as sum-frequency generation (SFG) or difference-frequency generation (DFG), typically driven by strong pump fields to facilitate spectral translation across hundreds of nanometers. Although numerous experimental demonstrations have been carried out in free-space configurations \cite{ates_two-photon_2012, de_greve_quantum-dot_2012, zaske_visible--telecom_2012} that offer excellent control of spatial modes and filtering capabilities, they often lack scalability and operational robustness, making them impractical for out-of-lab deployment. Recent efforts have aimed at developing integrated or fiber-compatible platforms to overcome these limitations. Chip-scale approaches using micro-ring resonators in GaAs or silicon photonics \cite{absil_wavelength_2000, jin_efficient_2013, Foster2007, li_efficient_2016} offer a high degree of miniaturization and potential CMOS compatibility. However, these systems tend to operate over narrow spectral windows, are sensitive to fabrication variations, and often suffer from limited external efficiency owing to the challenging fiber coupling and tight phase-matching constraints. For full system integration, it remains essential to focus on practical issues such as coupling losses, noise, and packaging. fiber-pigtailed waveguides, such as PPLN modules, offer a valuable compromise between efficiency and integration \cite{Murakami2023}, although the preservation of quantum properties and the overall system efficiency still leaves room for improvement.\\
In this study,  we focus on the conversion of photons emitted around $900~$nm~\cite{somaschi_near-optimal_2016, morrison_bright_2021, da_lio_pure_2022}, as this wavelength is at the crossroads of the most performing quantum dots (QD)~\cite{somaschi_near-optimal_2016} and other solid-state emitters, such as SiC~\cite{lukin_integrated_2020}, envisioning true single-photon emission and quantum repeater-based networks operating in the telecom range. We present a fiber-based QFC system that leverages fiber-to-chip coupling and achieves a high overall efficiency 48.4 \% from 925 to 1560 nm. It operates on a commercial fiber-pigtailed single photon source based on QDs \cite{margaria2024}. We demonstrate the full preservation of the unicity and indistinguishability of single photons produced by a QD, with $g^{(2)}(0)=0.044(1)/0.051(1)$ and indistinguishability = $79.3(3)\%/80.0(8)\% $ before and after conversion, respectively. Finally, by leveraging a tunable pump laser, we demonstrate the flexibility of the conversion process over a 10 nm range, enabling the alignment of the telecom wavelengths of converted single photons originating from different QDs, a common scenario in realistic quantum networks.

\section{experiment}

\subsection{Single-photon source and conversion modules}

\begin{figure*}[t]
    \centering
    \includegraphics[width=\linewidth]{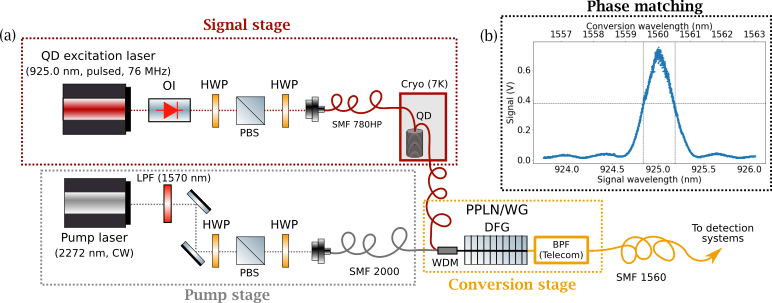}
    \caption{(a) Experimental setup of the frequency conversion interface. The signal and pump beams are going through the HWP and PBS to control the polarisation and power. The SMF are used to clean the spatial modes. Overlapping is performed using a WDM coupled into the PPLN/WG. The dashed lines correspond to free-space propagation. We filter the converted photons using telecom DWDM.
    (b) Phase matching of conversion process with $T_{PPLN} = 43.4$°C. FWMH are extracted, for the signal (FWHM$_{NIR} = 0.40 \pm 0.02$ nm) and the conversion (FWHM$_{conv} = 1.00 \pm 0.02$ nm).}
    \label{setup}
\end{figure*}

The experimental setup for the QFC is shown in Fig. ~\ref{setup}. The conversion interface can be divided into three stages. In the signal stage, a picosecond Ti:Sa laser (76 MHz repetition rate) is utilized to pump a fiber-pigtailed single-photon source based on a self-assembled InGaAs quantum dot embedded in an optical cavity. The polarisation and power of the pulses were meticulously controlled using half-wave plates (HWP) and a polarisation beam splitter (PBS). The QD is embedded within a micropillar structure composed of distributed Bragg reflectors fabricated using an advanced in situ lithography technique~\cite{senellart2017high}. A single-mode fiber (SMF) is precisely positioned on top of the micropillar at room temperature, with nanometer scale accuracy, maintained from 300 K to 7 K  \cite{margaria2024}. The complete system developed by \textit{Quandela} is fully operational and integrated. It comprises a pulse-shaping stage designed to ensure optimal matching between the cavity mode and spectral bandwidth of the excitation pulse, followed by injection into the fibered source maintained at 7 K within a closed-cycle cryostat. The source was operated with a 0.7 nm blue detuned laser exploiting the phonon-assisted excitation scheme \cite{thomas2021bright}. The final stage involves the collection of single photons at a wavelength of $\lambda_{SP}=925.7$ nm, with band-pass filters (BPF) employed to eliminate residual excitation laser light at $\lambda_{exc}=925.0$ nm.

The conversion stage is devoted to the frequency conversion process through difference-frequency generation (DFG). This nonlinear effect occurs within a 4 cm long periodically poled Lithium Niobate waveguide (PPLN/WG) with a poling period of $\Lambda = 25.45\,\mu m$. The waveguide is specifically designed with a poling period tailored for wavelength conversion from 925.7\,nm to 1560\,nm, ensuring compliance with both energy conservation and quasi-phase-matching conditions:
\begin{eqnarray}
\left\{
    \begin{array}{r c l}
     \omega_{sig} - \omega_{pump} & = & \omega_{conv}, \\\\
     \textbf{k}_{sig} - \textbf{k}_{pump} + \dfrac{2\pi}{\Lambda}\textbf{n} & = & \textbf{k}_{conv},
    \end{array}
    \right.
\end{eqnarray}
where $\Lambda$ is the poling period of the crystal and $\textbf{n}$ is a unitary vector.

A pump laser beam at 2272 nm is employed to satisfy the required conditions for frequency conversion and constituted the pump stage.  The laser operating in continuous wave mode (CW) is also injected into the SMF. Appropriate optical components are employed to control the polarisation and mitigate photonic noise. \\
The conversion stage is fully fiber-integrated. A 980/1560 nm wavelength demultiplexer is used to combine the signal and pump beams into the waveguide. For alignment, the output fiber is cleaved and positioned with a 3-axis micrometric mount placed less than 1 mm from the waveguide facet. Thanks to the large waveguide cross-section (12 µm × 12 µm), the near-perfect mode matching between the waveguide and the fundamental spatial mode of the signal, together with the anti-reflection coating, enables a high coupling efficiency (88\%) at the signal wavelength without requiring microlenses. In contrast, the pump beam exhibits a spatially multimode profile, which leads to suboptimal coupling into the WG (55\% efficiency). These injection losses are compensated by using a high pump power to reach maximal conversion efficiency. The converted photons are collected with a standard telecom fiber (SMF28) at 1560 nm, aligned using the same method as for the input. Coupling optimization is performed in the classical emission regime by replacing the SPS with a 925 nm CW laser. A filtering stage composed of four telecom dense-wavelength demultiplexers (DWDM) is used to achieve a high extinction of residual pump photons and non-converted single photons.

\subsection{Phase-matching}

The phase-matching bandwidth should exceed the spectral width of the photons, which is approximately $\Delta\lambda_{SP} \simeq 4$ pm,  to achieve complete conversion, but also cover a wider conversion range for more flexibility. To test our QFC, we perform measurements in the classical regime by substituting the QD with a CW laser tuned to approximately 925 nm. The primary tunable parameter that influences the phase-matching width is crystal temperature. The PPLN crystal is mounted on an oven connected to a temperature controller, which allows for precise stabilisation and temperature adjustment.  A linear drift is performed, and the converted signal is detected using a photodiode. Fig.~\ref{setup}(b) shows the resulting signal, from which the full width at half maximum (FWHM) of the central lobe is extracted. We achieve FWHM$_{NIR} = 0.40 \pm 0.02$ nm at $\lambda_{NIR} = 925.0$ nm at a temperature of $T_{PPLN} = 43.4$°C, satisfying the condition of FWHM$_{NIR} \gg \Delta\lambda_{SP}$. 

\begin{figure}
    \centering
    \includegraphics[width=\linewidth]{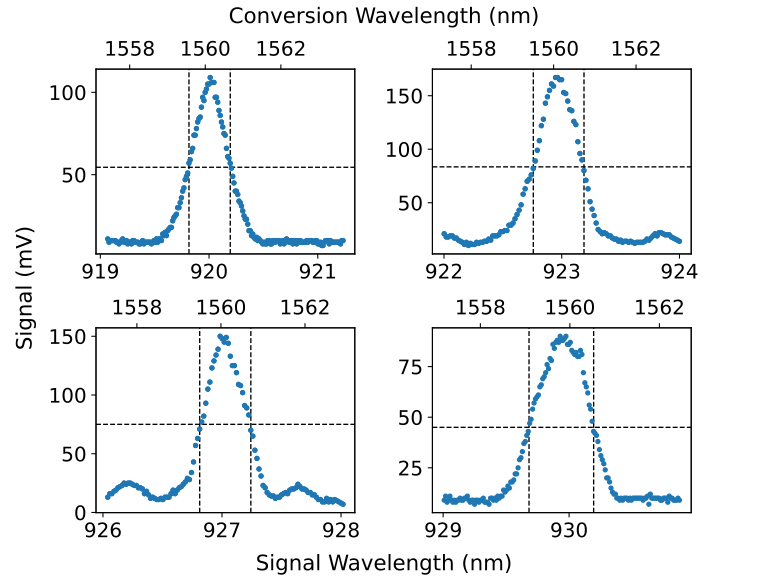}
   \caption{Phase matching curves with different central signal wavelength. The temperature is adjusted to conserve the central conversion wavelength of 1560 nm. For these four curves: 0.37 nm $<$ FWMH$_{NIR}$ $<$ 0.49 nm, with the PPLN temperature from 33.0 °C to 82.5 °C. }
\label{flex}
\end{figure}

To further optimise the system, we conduct phase-matching measurements to ensure that our conversion interface could accommodate wavelength mismatches from different single-photon sources. By adjusting the central wavelengths of the pump beam and modifying the temperature of the PPLN crystal, we maintain phase-matching centred at 1560 nm. The phase-matching bandwidth remains the same while varying the wavelength of the signal photons from 920 to 930 nm (Fig. ~\ref{flex} a-d), with FWHM$_{NIR}$ between $0.37$ nm and $0.49$ nm, which are still lower than the spectral width of the photons. The crystal used in this setup features six waveguides with different poling periods, extending the spectral tuning range to approximately 60 nm. This range ensures the compatibility of the interface with various single-photon source technologies.

\subsection{Efficiency}

\begin{figure}[h!]
    \centering
    \includegraphics[width=\linewidth]{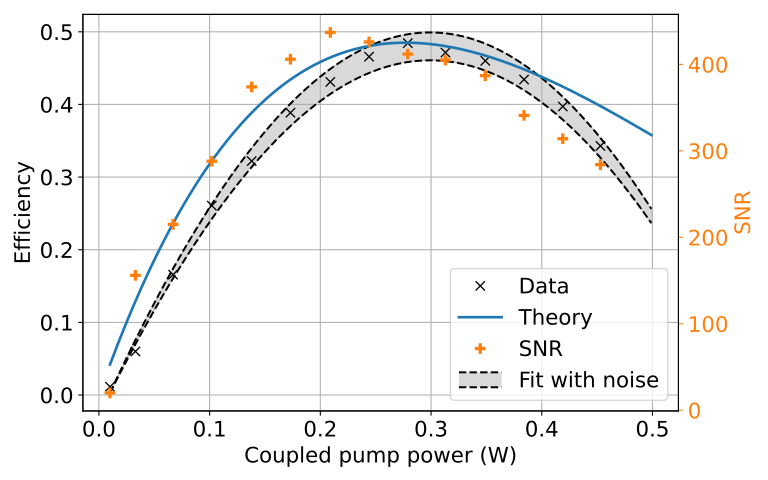}
    \caption{External conversion efficiency as a function of coupled pump power. The data (black) are fitted (grey zone) and compared with the theoretical curve (blue). Orange dots : Signal-to-noise ratio of the interface, depending on the coupled pump power.}
    \label{eff}
\end{figure}

The conversion efficiency is a key factor in evaluating the performance of the QFC interface. To determine the maximum achievable efficiency of a single photon from the QD, we measure the performance for various levels of coupled pump power in the waveguide. A superconducting nanowire single-photon detector (SNSPD) operating in the telecom band is employed to detect the number of converted photons at the output of the QFC. The external conversion efficiency, $\eta_{ext}$, is calculated by comparing the number of signal photons injected into the conversion interface, $N^{in}_{sig}$, with the number of converted photons at the output, $N^{out}_{conv}(P_{pump})$, as a function of the pump power:

\begin{eqnarray} 
\eta_{ext} = \frac{N^{out}_{conv}(P_{pump})}{N^{in}_{sig}}.
\end{eqnarray}
Fig.~\ref{eff} shows our results compared with the theory described by the following formula: 
\begin{eqnarray}
\eta(P_{pump}) & = & \eta_{max}sin^2\left(L\sqrt{\eta_nP_{pump}}\right),
\end{eqnarray}
where $\eta_{max}$, $P_{pump}$, $L$ and $\eta_n$ are the maximum conversion efficiency, coupled pump power, length of the nonlinear media, and normalised efficiency specific to the device, respectively.

The theoretical efficiency of DFG is subject to an optimal value, as excessive pump power can lead to reconversion processes that limit the efficiency. We achieve a maximum external conversion efficiency of $\eta_{ext,max}$ = 48.4 \% with a coupled pump power of $P_{pump}$ = 285\,mW. 

At then input of the QFC stage, the in-fiber single-photon rate is 2.8 MHz, corresponding to 3.7 \% brightness. After frequency conversion at the maximum efficiency, the converted photon rate is 1.3 MHz. The discrepancy between the experimental data and theoretical curve is attributed to the noise generated during the conversion process, primarily from two sources: Raman scattering and residual pump light. The filtering stage has been optimized to maintain high brightness while minimising the noise contribution. To assess noise performance, we measure the signal-to-noise ratio (SNR) by comparing detections inside and outside the phase-matching range of the conversion. In the optimal regime ($P_{pump}$ = 285\,mW), the SNR exceeds 400, demonstrating the effective noise suppression and high performance of the conversion interface.

\subsection{Unicity \& indistinguishability}

\begin{figure*}[t]
    \centering
    \begin{tabular}[c]{|c|c|c|}
        \hline
         & \textbf{Before QFC (925.7 nm)} & \textbf{After QFC (1560 nm)} \\
        \hline
            $\boldsymbol{g^{(2)}(0)}$ \\[-2.8ex]
            \hspace{1em}\includegraphics[width=0.2\linewidth]{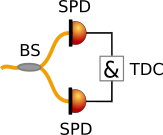}\hspace{1em}
          & \includegraphics[width=0.3\linewidth]{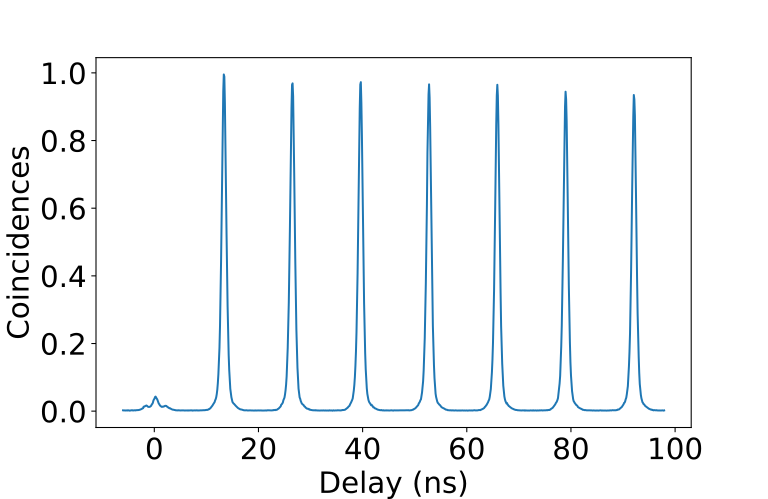} & \includegraphics[width=0.3\linewidth]{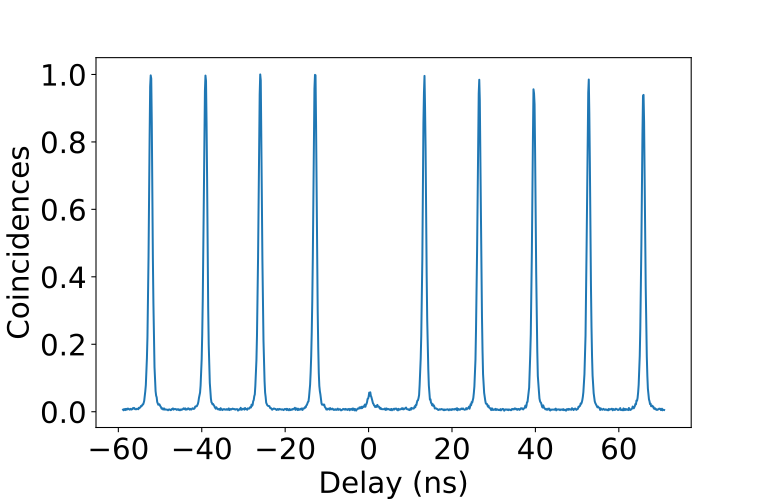} \\
        \hline
            $\boldsymbol{V_{HOM}}$ \\[0ex]
            \includegraphics[width=0.4\linewidth]{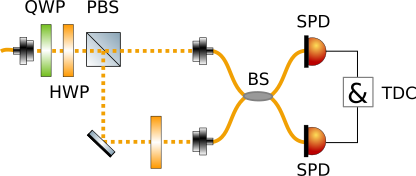} & \includegraphics[width=0.3\linewidth]{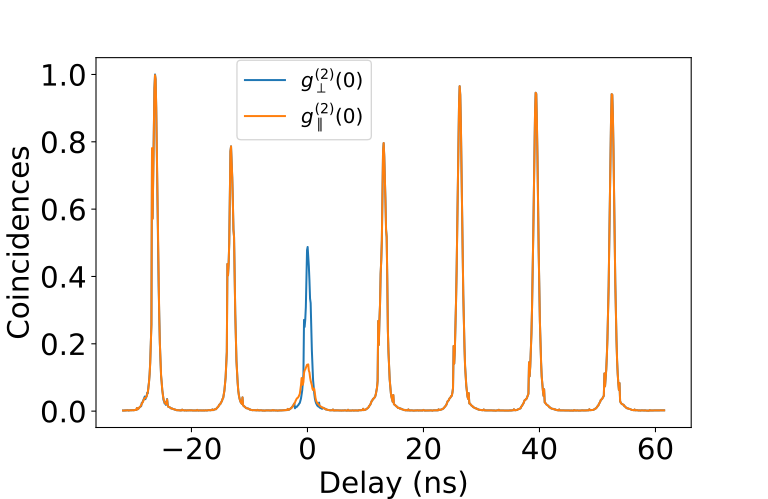} & \includegraphics[width=0.3\linewidth]{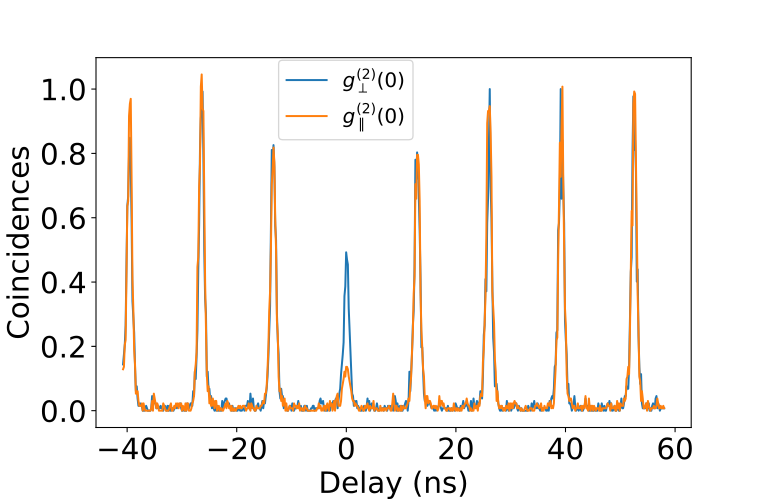} \\
        \hline
    \end{tabular}
    \caption{Characterization of single photon properties conservation. \textit{Left column} : Detection setups. $g^{(2)}(0)$ is measured using an HBT interferometer, recording coincidences on a time-to-digital converter (TDC) and receiving signals from single-photon detectors (SPD). The $V_{HOM}$ is extracted by using an UMZI, using a HWP to configure indistinguishable case (co-polarization) and distinguishable case (cross-polarization). \textit{Middle column} : Results in NIR band, at the output of the QD. \textit{Right column} : Results after conversion, at $1560$ nm.}
    \label{D}
\end{figure*}

We now compare the properties of the single photons before and after passing through the conversion interface. The initial characterisation focuses on the unicity (or single-photon purity) of the source and quantifies the probability of emitting more than one photon per pulse. The relevant parameter is the second-order autocorrelation coefficient at zero delay $g^{(2)}(0)$, which is defined as 

\begin{equation}
    g^{(2)}(0) = \dfrac{\left\langle \hat{n}(\hat{n}-1) \right\rangle}{\left\langle \hat{n} \right\rangle^2},
\end{equation}
where $\hat{n}$ denotes the photon-number operator. 
To measure this parameter, we utilize two Hanbury-Brown and Twiss (HBT) interferometers, one in the NIR band and the other in the telecom band. Each interferometer is configured with a 50/50 beam splitter (BS), two detectors, and a time tagger to record coincidence events (Fig.~\ref{D}). Coincidence histograms are generated as a function of the delay between detectors. To extract the values of $g^{(2)}(0)$, we compare the area of the zero-delay peak with the mean area of the other peaks, which correspond to correlations between successive photons. This comparison is made using the following formula.

\begin{equation}
    g^{(2)}(0) = \dfrac{A_0}{\dfrac{1}{N}\overset{N}{\underset{n\ne 0}{\sum}}A_n},
\end{equation}

with $A_{0,...,n}$ peak area at delays $0,...,n$ and $N$ total peak numbers. Histograms are shown in the upper row of Fig.\ref{D}, highlighting a reduction in coincidences at zero delay. We measure $g^{(2)}(0)_{NIR} = 0.044(1)$ and $g^{(2)}(0)_{Tel} = 0.051(1)$ before and after the conversion, respectively, indicating good preservation of the single-photon purity. Note that the single-photon purity in this experiment was probably limited by the non-optimal laser filtering stage. A higher single-photon purity can be achieved with $g^{(2)}(0)$ values below $10^{-2}$ with optimised excitation schemes and filtering~\cite{arakawa2020progress}.

The second step involves characterising the indistinguishability between single photons consecutively emitted by the QD. Indistinguishability here means that the photons are identical in all modes, including polarisation, temporal, spatial, and spectral modes. To quantify this, we conduct a Hong-Ou-Mandel (HOM) experiment involving two-photon interference between consecutive photons. In the HOM experiment, two indistinguishable photons are sent into a 50/50 beam splitter (BS), leading to a dip in the coincidence histogram at zero delay, known as the HOM dip. The visibility of this interference, $V_{HOM}$, provides a measure of the indistinguishability of photons emitted by the QD. To demonstrate this property, we set up an unbalanced Mach-Zehnder interferometer (UMZI) with a relative delay corresponding to the repetition rate of the QD excitation laser (76 MHz, equivalent to 13 ns). Two UMZIs are constructed, one for the NIR band and one for the telecom band, with a free-space section to finely adjust the relative delay. The photons are prepared in a diagonal polarisation state and separated by a polarisation beam splitter (PBS) at the input of the interferometer. SNSPDs are used to detect photons and coincidence events are recorded using a time-to-digital converter (TDC). To evaluate $V_{HOM}$, we compare the coincidence counts when the photons are maximally distinguishable and when they are indistinguishable. The polarisation mode is adjusted using an HWP to prepare photons with either co- or cross-polarisation. The visibility is extracted using the following formula:

\begin{equation}
    V_{HOM} = 1 - \dfrac{g^{(2)}_\parallel (0)}{g^{(2)}_\perp (0)},
\end{equation}
with $g^{(2)}_{\parallel (\perp)} (0)$ zero-delay coincidence peak for the indistinguishable (distinguishable) case. This raw value is corrected to obtain the single-photon indistinguishability ($M_s$), correcting for accidental coincidences caused by the nonzero $g^{(2)}(0)$ of the source~\cite{ollivier2021hong}: 
\begin{equation}
    M_s = \dfrac{V_{HOM} + g^{(2)}(0)}{1 - g^{(2)}(0)}.
\end{equation}
We compare the results before and after the conversion interface by measuring the areas of the zero-delay peaks. For the raw data, we obtain $V_{HOM}^{NIR} = 71.4(3)\%$ and $V_{HOM}^{Tel} = 70.8(8)\%$, indicating ideal preservation of indistinguishability. After applying the corrections, the indistinguishability values are $M_s^{NIR} = 79.3(3)\%$ and $M_s^{Tel} = 80.0(8)\%$ before and after conversion, respectively. The non-ideal indistinguishability of the source under study arises from the low Purcell factor experienced by the source in the low-quality factor cavity. These results underscore the effective coherence preservation achieved using the conversion interface.

\section{Discussion}

\begin{table*}
    \centering
    \begin{tabular}{|c|c|c|c|c|c|c|}  
        \hline
        Work & $\lambda_{NIR}$ & QFC $\eta_{max,ext}$ & $g^{(2)}(0)$ NIR/TC & $M_s$ NIR/TC & SP rate NIR/TC & Inj./Coll. QFC \\
        \hline
        \hline
        \textbf{This work} & \textbf{925 nm} & \textbf{$\boldsymbol{48.4~\%}$} & \textbf{$\boldsymbol{4.4/5.1~\%}$} & \textbf{$\boldsymbol{79.3/80.0~\%}$} & \textbf{$\boldsymbol{2.8/1.3}$ MHz} & \textbf{fully fiber-coupled} \\
        \hline
        Da Lio et al.~\cite{da_lio_pure_2022} & 945 nm & $40.8~\%$ & $2.0$/$2.4~\%$ & $93.5$/$94.8~\%$ & 2.21/0.905 MHz & partially fiber-coupled \\
        \hline
        Morrison et al.~\cite{morrison_bright_2021} & 942 nm & $35~\%$ & $4.0$/$4.3~\%$ & $95$/$67~\%$ & 1.46/0.456 MHz & partially fiber-coupled \\
        \hline
        Rickert et al.~\cite{Rickert2024} & 1550 nm & No conversion & -/$<4~\%$ & -/$79$-$(55)~\%$ & -/$0.6$-$(1.2)$ MHz & fully fiber-coupled \\
        \hline
    \end{tabular}
    \caption{Comparison of this work with similar QFC at the state-of-the-art. Efficiencies are reported fiber-to-fiber. The coupling scheme indicates whether the PPLN stage is fully fiber-coupled or includes a free-space interface. The last line represents the current state-of-the-art in the direct generation of a telecom single-photon source with QDs~\cite{Rickert2024}. Two numbers are those reported, the higher (lower) indistinguishability being reported for half (full) brightness.}
    \label{artconv}
\end{table*}

The characterization of the QFC interface is summarized in Table~\ref{artconv}. We now compare our results with state-of-the-art interfaces based on the DFG process for single-photon conversion from the NIR to telecom band. We first highlight the performance of our QFC independently of the SPS, because our QFC can be adapted to any SPS. Our QFC demonstrates the highest end-to-end conversion efficiency of approximately 900 nm while preserving the unicity and indistinguishability properties of the SPS. Additionally, the spectral tunability of our scheme, which leverages the PPLN crystal and pump laser, offers significant flexibility for interfacing with a multitude of NIR SPS, allowing for the integration of dissimilar SPS. A key distinction of our approach is the advantageous use of optical fiber components throughout the interface. Single photons are collected in a fiber and transmitted to a fully fibered conversion stage, maintaining a high coupling efficiency and generating minimal losses. The fiber-based system provides notable compactness, with the pump preparation, conversion, and filtering stages mounted on a 400 × 400 mm breadboard, enhancing transportability and making the setup suitable for real-world applications. While reference \cite{murphy_dataset_2024} reports a higher internal efficiency, their setup involves a more complex fiber connection, which poses challenges for integration with standard SMF28 fibers or other common systems. This complexity may hinder practical deployment and integration compared with our more streamlined and transportable fiber-based approach.\\

Table~\ref{artconv} further highlights that the single-photon source performance obtained at the end of the QFC system represents a new record  value in terms of the photon rate. We note that this is achieved for the very first time using a commercial fiber-pigtailed source, where recent progress has shown a factor of six increase in fibered brightness in the NIR~\cite{margaria2024}. The direct generation of single photons in the telecom band from QDs has attracted significant interest in recent years~\cite{pfister_telecom_2024, murphy_tunable_2024, murakami_low-noise_2024, Rickert2024, alqedra_indistinguishable_2025, pettit_monolithically_2025, hauser_deterministic_2025, costa_telecom_2025}, achieving a promising performance in terms of unicity and indistinguishability. Notably, the recent work in~\cite{hauser_deterministic_2025} which we include in Table~\ref{artconv}, illustrates  significant advances in these metrics with, for the first time, a photon indistinguishability of 79~\% at 0.6 MHz. These values represent important progress, approaching for the first time the performance of the best NIR SPS converted to the telecom band. However, we note that the QFC is an important asset for bringing the capability of NIR QD sources to generate spin-photon entanglement~\cite{cogan2023deterministic, coste2023high, huet2025deterministic} into the telecom range for quantum networks and distributed optical quantum computing.\\

Improvements to the QFC interface are envisioned to enhance performance and operational usage. The principal source of loss is the filtering stage after conversion, which corresponds to $-0.8$ dB owing to several DWDMs connected in series. A strong pump beam can be better suppressed with custom fiber optics featuring a high extinction ratio tailored to our specific wavelengths. Additionally, the non-ideal coupling of the signal with cleaved fibers in the waveguide (currently at 88\%) can be improved through pigtailing techniques. Furthermore, the internal efficiency of the PPLN has been characterized in the classical regime to estimate the total achievable efficiency by measuring $\eta_{int} \sim 66\%$. The non-unity internal efficiency is mainly due to the spatial mode mismatch between the signal and the fundamental mode of the waveguide. QFC developed using Lithium Niobate on insulator (LNOI) have achieved higher internal efficiencies and hold promise for integration through nanophotonics chips~\cite{wang2023quantum}.

Thus, our fiber-based interface offers an optimal combination of high efficiency, compactness, scalability, and quantum fidelity, which are key components for deploying QFC technologies in practical quantum information processing architectures.\\

In conclusion, we demonstrated an efficient and coherent QFC interface compatible with the best-performing SPS, such as the NIR QD. We achieved a state-of-the-art end-to-end efficiency ($\eta_{ext} = 48.4 \%$). We also report the preservation of unicity and indistinguishability, which are key parameters for the development of quantum information technologies. The fibered conversion stage ensures the compactness and transportability of the system, offering viable solutions for real field and deployed applications. This work paves the way for quantum key distribution (QKD) solutions, implementing one-way links based on single photons as qubit carriers~\cite{morrison2023single} and extending to long-range networks to achieve quantum repeaters based on quantum state teleportation outside laboratory settings.

\section*{Acknowledgment}

This work was conducted within the framework of the OPTIMAL project granted by the European Union by means of the Fond Européen de développement regional (FEDER). The authors also acknowledge the financial support from the Agence Nationale de la Recherche (ANR) through the projects LIGHT (18-ASTR-0024), the European Comission for the Quantera project (INQURE ANR-22-QUA1-0002) and the French government through its Investments for the Future program under the Université Côte d'Azur UCA-JEDI  project  (Quantum@UCA)  managed  by  the  ANR (ANR-15-IDEX-01).

\section*{Author information}

\section*{Competing interests}
P. Senellart and N. Somaschi are co-founders, and the CEO and CSO of Quandela, respectively.

\section*{Data Availability}
The data are available from the authors upon request.

\end{document}